\documentstyle[aps,preprint]{revtex}
\begin{document}

\draft
\title{Critical behavior  of the $S=1/2$ Heisenberg ferromagnet:
 \\A Handscomb quantum Monte Carlo study} \vspace{0.5cm}

\author{Adauto J. F. de Souza}\address{Departamento de F\'{\i}sica e 
Matem\'atica, Universidade
Federal Rural de Pernambuco, 52171-030 Recife PE, Brazil \\ 
Departamento de F\'{\i}sica, Universidade Federal
 de Pernambuco, 50607-901 Recife - PE, Brazil}
\vspace{0.5cm}
\author{U. M. S. Costa}\address{Departamento de F\'{\i}sica, Universidade
 Federal do Cear\'a, 60470-455 Fortaleza - CE,
Brazil}
\vspace{0.5cm}
\author{M. L. Lyra}\address{Departamento de F\'{\i}sica, 
Universidade Federal
 de Alagoas, 57072-970 Macei\'o - AL, Brazil}

\maketitle
\vspace{1.0cm}

\begin{abstract}
We investigate the critical relaxational dynamics of the  $S=1/2$ Heisenberg 
ferromagnet on a simple cubic lattice within the Handscomb  
prescription on which it is a diagrammatic series expansion of 
the partition function that is computed by means of a Monte Carlo 
procedure. Using a phenomenological renormalization group analysis 
of graph quantities related to the spin susceptibility and order 
parameter, we obtain precise estimates for the critical exponents 
relations $\gamma / \nu = 1.98\pm 0.01 $ and
$\beta /\nu = 0.512 \pm 0.002$ and for the Curie temperature 
$k_BT_c/J = 1.6778 \pm 0.0002$. The critical
correlation time $\tau_{int}$ of  both energy and susceptibility
is also computed.
We found that the number of Monte Carlo steps needed to 
generate uncorrelated diagram configurations scales with 
the system's volume. We estimate the efficiency of the 
Handscomb method comparing its ability in dealing with
the critical slowing down  with that of other quantum
and classical Monte Carlo prescriptions.\\

\end{abstract}

\newpage

\section{introduction}

The slowing down of the relaxation to thermal equilibrium 
is an important physical phenomenon associated to the build
 up of long-range correlations at the critical point of spin 
systems. According to the finite size scaling hypothesis, the
 critical relaxation time scales with the system size as 
$\tau\propto L^z$, with the dynamical critical exponent $z$ 
governing the rate of convergence towards thermal equilibrium. 
The value of $z$ depends on the particular equation of motion 
of the order parameter and of the conservation laws that apply 
to the spin system. In particular, the dynamic scaling theory 
predicts that for the classical isotropic Heisenberg model with 
conserved order parameter $z= d-\beta /\nu$ ($z\sim 2.5$ for 
$d=3$)\cite{Halperin}, which is consistent with both
 experiments\cite{Boni1,Boni2} and numerical 
simulations\cite{Chen}. For a recent review see Ref.~\cite{Landau}

Monte Carlo simulation is a powerful tool to study the equilibrium
 properties of a spin system  through some stochastic  relaxational
 dynamics. It defines a Markovian process in which the associated stochastic
model evolves in the phase space according to certain transition
probabilities. The transition rates between two
 distinct spin configurations are imposed to satisfy the detailed 
balance condition in order to lead the system to its equilibrium state
 distribution. The dynamical evolution of the stochastic model
can be thought of resulting from the interactions among  its many
degrees of freedom.
 However, the evolution of the model-system under the
Monte Carlo dynamics do not need to be consistent with any
motion equation.  In this context,
the critical exponent $z$ characterizes
therefore the rate at which a particular set of stochastic dynamical
 rules generates uncorrelated spin configurations.

 A general feature of Monte Carlo simulations of classical spin models 
is that cluster dynamics are usually much more effective in overcoming the 
critical slowing down than those which are based on single spin-flip
procedures. For example, the cluster Monte Carlo algorithm introduced by 
Swendsen-Wang\cite{Swendsen} is known to have a very fast convergence to 
equilibrium at the critical point in contrast to the slower convergence 
of local dynamics such as Metropolis and heat-bath. For the isotropic 
classical Heisenberg model it has been found that $z\sim 2$  for 
Metropolis\cite{Peczak} and $z\sim 0$  for the single cluster\cite{Holm1,Holm}
 Monte Carlo dynamics.

The traditional spin-flip Monte Carlo algorithms as applied to 
classical spin systems
can not be directly extended to quantum spin models. The problem 
resides in the fact
that the Hamiltonian is not, in general, diagonal in the spin configuration 
basis. A quantum  Monte Carlo
 method was first introduced by Handscomb
 for the calculation of the thermodynamical properties of quantum
Heisenberg ferromagnets\cite{Handscomb}. 
 The main difference of this technique
in comparison to  traditional Monte Carlo algorithms is that
 the sample space is not related to
any kind of physical phase space. It is the diagrammatic series
 of the partition
function that is calculated by means of the Monte Carlo method. 
 The Handscomb method has been successfully used to compute the 
thermodynamical properties of the Heisenberg $S=1/2$
ferromagnet\cite{Lyklema,Lee,Rojdestvenski1,Rojdestvenski2} and
extended to a form applicable to a series of quantum spin 
models\cite{Sandvik}.

Another algorithm  commonly used in Monte Carlo simulations of quantum
 spin models is based on the use of the generalized Trotter formula to
 map the quantum system onto a classical system with an additional 
imaginary time dimension\cite{Suzuki}. The $S=1/2$ Heisenberg model has
 been extensively studied within this line. Recently, a decoupled cell 
method for quantum Monte Carlo based on the Suzuki-Trotter approach has 
been used to compute the critical dynamical exponent $z$ of the $S=1/2$ 
Heisenberg model on the simple cubic lattice\cite{Sisson}. It was found 
that $z\sim 2$ which is quite similar to the value obtained from simulations
 of the classical Heisenberg model under Metropolis dynamics\cite{Peczak}.

In the present work, we are going to investigate  both the static and dynamic critical
properties of
the $S=1/2$ Heisenberg ferromagnet on the simple cubic lattice   by means 
of the Handscomb dynamics. We will employ a phenomenological 
renormalization group
to obtain precise estimates of the critical temperature and 
some static critical
exponents. By employing a so called
moving block bootstrap (MBB) technique\cite{mbb}, we are going to calculate the
 equilibrium relaxation time for the energy and susceptibility
at criticality.
The critical time-displaced equilibrium correlation function will be
computed as well. We will employ a finite size scaling analysis
of the  equilibrium relaxation time to obtain the critical dynamical
 exponent $z$ associated to the Handscomb dynamics and we contrast it with the
 results from the quantum decoupled cell method and the results from  distinct
 Monte Carlo simulations of the classical Heisenberg model.

\section{The Handscomb Monte Carlo method}

Let us briefly draw the main ideas of the Handscomb method. Consider
 the Hamiltonian of a quantum spin system to be given by
\begin{equation}
H=\sum_i^{N_0} H_i ,~~~[H_i,H_j]\neq 0
\end{equation}
The canonical average of a physical observable $A$ can be
expanded in the form
\begin{equation}
\left< A \right> = \frac{Tr[Aexp(-\beta H)]}{Tr[exp(-\beta H)]} =
\sum_r\sum_{C_r}A(C_r)p(C_r)
\end{equation}
where $\beta = 1/k_BT$, $\sum_{C_r}$ denotes a summation over
all ordered  sets of indices $C_r\equiv \{i_1,i_2,...,i_r\}$
(Mayer diagrams) and
\begin{equation}\begin{array}{l}
A(C_r)\equiv\frac{Tr[AH_{i_{1}}....H_{i{r}}]}{Tr[H_{i_{1}}....H_{i_{r}}]}\\[3mm]
p(C_r)\equiv\frac{\frac{(-\beta)^r}{r!}Tr[H_{i_{1}}....H_{i_{r}}]}
{\sum_r\sum_{C_{r}}\frac{(-\beta)^r}{r!}Tr[H_{i_{1}}....H_{i_{r}}]}
\end{array}
\end{equation}
Once $p(C_r)\geq 0$, it can be considered as a probability
distribution and the canonical averages  can be written as $\left< A\right> =
\left< A(C_r)\right>_{p(C_r)}$. This is the case of the Heisenberg $S=1/2$ ferromagnet.
 The Hamiltonian can be
represented in terms of transposition operators as
\begin{equation}
H = -J\sum E_{i,j}
\end{equation}
so that the relevant trace to be computed is that of a permutation
operator
\begin{equation}
Tr P(C_r)\equiv Tr [E_{(i,j)_1}E_{(i,j)_2}...E_{(i,j)_r}] =
2^{k(C_r)}
\end{equation}
where $k(C_r)$ is the number of cycles in the irreducible
representation of the permutation $P(C_r)$. It is straightforward to show that any
 physical observable can be expressed in terms of the diagram structure.
For instance, the internal energy is related to the average number of
transposition operators in the diagrams and the susceptibility to the average
size of the cycles in the diagram's irreducible representation \cite{kalos}.

The Handscomb Monte Carlo method organizes a
random walk in the space of the diagrams $C_r$ which has $p(C_r)$
as the limit distribution. The dynamics suggested by Handscomb
consists
of three types of steps: (i) Step forward, chosen with probability $f_r$,
which tries to include a
randomly chosen bond to the right of the permutation operator;
(ii) Step backwards, chosen with probability $1-f_r$, which tries
to
remove a bond from the left of $P(C_r)$; (iii) cyclic
transposition, chosen when step backwards is rejected, which
moves a bond from the left to the right. The transition
probabilities for performing each movement on the space of Mayer's
diagrams are chosen in order to satisfy the detailed balance
condition.

After a single step of the Handscomb Monte Carlo dynamics, the irreducible 
representation of the sequency $C_r$ can have its cycle structure changed 
considerably. When two sites belong to  distinct cycles, the insertion of 
the corresponding bond  results in the coalescence of the two  cycles of 
permutations. On the other hand, i.e., when the sites belong to the same
 cycle, the insertion breaks the cycle in two new ones. The same process 
occurs when a bond is removed from the sequence. Therefore, entire sets of 
sites can have their status changed during a single Monte Carlo step and, in 
this sense, the Handscomb dynamics is similar to the classical Monte Carlo 
cluster algorithms. 

\section{Finite size scaling for the susceptibility and order parameter}

The susceptibility per spin of the quantum $S=1/2$ Heisenberg model is written 
in terms of the cyclic structure of the irreducible representation of $C_r$ as\cite{kalos}
\begin{equation}
\beta\chi = \frac{1}{N}\left<\sum_{j=1}^{k(C_r)}a_j^2\right> _P
\end{equation}
where $a_j$ is the length of the $j$-th cycle of the permutation $P(C_r)$ and
 $\langle \cdots \rangle_P$ denotes an average with respect to the $C_r$-space probability
distribution. In figure 1, we show our results for the susceptibility from 
lattices of  $L^3$ spins with $L= 16, 24$ and, $32$. In these simulations
$150L^3$ Monte Carlo steps (insertion, removal or bond permutation) were enough to let
the system evolve to an equilibrium diagram configuration starting from an
 initial diagram containing no transpositions. After equilibrium was reached,
 we averaged over $2\times 10^4$ distinct diagrams, $10^3$
 MCS apart. These results were averaged over $10$ distinct realizations of the numerical
 experiment. The susceptibility exhibits a critical behavior around
 $k_BT_c/J\simeq 1.68$ in agreement with previous Monte Carlo estimates\cite{Chen1}.
 For $T > T_c$, $\chi$ is only weakly dependent on the
system size; whereas it is nearly proportional to $L^3$ at low
temperatures. Notice that $\chi$ equals the magnetization second moment for
temperatures below $T_c$ once the magnetization is strictly zero within the 
Handscomb prescription.  

 In order to obtain a precise estimate of the critical temperature,
 we implemented a  phenomenological
 renormalization group analysis of the data from finite size lattices
 as introduced by Nightingale\cite{Nightingale}. The basic assumption is 
that near the transition
 the susceptibility of a finite lattice of linear size $L$ scales as
 \begin{equation}
 \chi (T,L) = L^{\gamma/\nu}f_{\pm}(tL^{1/\nu})
 \end{equation}
where $t=|(T-T_c)/T_c|$ and ($\pm$) stands for distinct scaling
functions above and below $T_c$. The renormalization of temperature
 is defined by the following transformation relating lattices of two different sizes,
 $L$ and $L^{\prime}$
 \begin{equation}
 \chi (T,L) = (L/L')^{\gamma/\nu}\chi (T',L')
 \end{equation}
 with the fixed point giving  $T_c$. Then a set
 of auxiliary functions is introduced as
 \begin{equation}
 g_{\chi}(T,L,L') = \frac{\ln{[\chi (T,L)/\chi (T,L')]}}{\ln{(L/L')}}
 \end{equation}
 and these intercept as a function of temperature at a common
 point from which we can directly measure $T_c$ and $\gamma/\nu =
 g_{\chi}(T_c,L,L')$. In figure 2 we plot the auxiliary functions
 $g_{\chi}(T,L,L')$ for typical renormalizations.
 Using all possible renormalizations with lattice sizes $L=16, 24, 32, 40$ 
and $48$,  we estimate $k_BT_c/J= 1.677 \pm 0.001$ and $\gamma/\nu = 1.98 \pm 0.01$.
These values are one order of magnitude more accurate than the previous Monte Carlo estimates 
 from simulations on small lattices ($L\leq 10$) which reported
$k_BT_c/J = 1.68 \pm 0.01$\cite{Chen1}.

An even more accurate value for the critical 
temperature can be found by employing 
a renormalization study of critical quantities 
which are known to depict smaller
 fluctuations near the critical point such as 
the magnetization itself. Unfortunately,
 as we  mention before,  the magnetization is exactly zero for  all temperatures due
 to an intrinsic symmetry of the Handscomb dynamics. However, we can explore the  
cycle structure of the Mayer diagrams to introduce a graph quantity which  display
 the same critical behavior of the order parameter. In the simulations of classical
 spin models, such a quantity is the size of the largest cluster of spins which are 
in the same state. This might suggest that the largest cycle
within a diagram in the context of  Handscomb MC,
may exhibit the same scaling behavior as the magnetization.
Therefore, we will introduce a graph order parameter as the average
size of the largest cycle of permutations. 

In figure 3, we plot the average size 
of the largest cycle (normalized by the total number of sites) as a function
 of temperature from simulations on lattices
 with $L=16, 24$ and, $48$.  From this figure, one can see that the  average size 
of the largest cycle depicts an
overall behavior similar to the one expected for an order parameter, and it will be 
considered as a true order parameter
so forth. It also indicates a phase transition around $k_BT_c/J \approx 1.68$.
 A renormalization analysis performed on  the order
 parameter data is shown in figure 4. From these data we found $k_BT_c/J = 1.6778 \pm 0.0002$
and $\beta /\nu = 0.512 \pm 0.002$. From the best of our knowledge, the presently 
reported values for $k_BT_c/J$, $\gamma /\nu$ and $\beta /\nu$ are the most accurate
Monte Carlo estimates up to date for the quantum 3D Heisenberg ferromagnet.
Our quoted value for $T_c$ is in complete agreement with the most accurate
high-temperature series study which yielded $J/k_BT_c = 0.5960(5)$~\cite{oitmaa}.
The critical exponents
are in excellent agreement with the best estimates for the classical Heisenberg
 ferromagnet\cite{dplandau93}.

\section{Critical Relaxation of the Spin 1/2 Heisenberg Model}

The critical relaxation within the Handscomb prescription can be investigated
 by computing some equilibrium time-displaced correlation functions $C(t)$ 
at the Curie temperature.  We look at the equilibrium relaxation time
$\tau$ which is expected to depict a power-law
increase with the system size $L$ whose exponent characterizes the critical 
relaxation process. In particular, it governs the size dependence of the rate at which
 uncorrelated configurations are generated during the Monte Carlo temporal 
evolution in phase space.

The fast growth  of the relaxation time is referred to  as the critical slowing down
which may be governed by several relaxation modes\cite{dplandau91}. One generally is interested
in the slower relaxation modes, i. e., the longest relaxation times. Therefore, it is
safer to work with the integrated correlation time given by
\begin{equation}
\tau_{int} = \int_{0}^{\infty} C(t) dt.
\end{equation}
In order to estimate $\tau_{int}$, we perform very long MC simulation on $L^3$
simple cubic lattices, with $L=16, 20, 24, 28, 32, 36, 40, 44, 48$,
 at the previously
calculated critical temperature $k_BT_c/J = 1.6778$.
The simulation started from a diagram containing no transposition and we observed
that typically $150 L^3$ configurations were needed to bring the system to equilibrium.
So we discarded the   appropriate number of configurations for equilibration, after
which we recorded the
susceptibility and
energy every $\delta t = 2000$ MCS, generating long
 equilibrium time series 
of $10^6$ measurements each.

The time-displaced correlation functions were obtained by 
$C_q(t) = \Delta_q(t)/ \Delta_q(0)$,
where $\Delta_q(t)$ is the autocovariance function given by
\begin{equation}
  \Delta_q(t) = \frac{1}{n-t} \sum_{i=1}^{n-t} 
(q_i - \langle q \rangle) (q_{i+t} - \langle q \rangle),
\end{equation}
$n$ is the length of the time series, and $q$ represents the 
physical property one is interested in.

Here, we computed $C_{\chi}(t)$ and $C_E(t)$, the correlation function of the 
susceptibility and energy respectively. Typical equilibrium traces of the 
susceptibility and energy are shown in figure 5, where the microscopic time 
scale used equals to $2000$ MCS. From these we can infer that the number of
MCS needed to generate two diagram configurations with uncorrelated  
susceptibilities is much smaller than the one required to generate
uncorrelated energies.

In practice, $\tau_{int}$ was estimated by
\begin{equation}
\tau_{int} = \sum_{t=0} C_q(t)
\end{equation}
and the sum was cut off at the first negative value of $C(t)$.

Despite of our long runs we were not able to get reliable estimates of $\tau_{int}$
by integrating $C(t)$
for the largest lattices simulated. It is well known  that $C(t)$
fluctuates wildly for large $t$, hampering the convergence of its integral.

On the other hand, in the context of MC simulations the error associated to a given quantity
can be written as\cite{binder}
\begin{equation}
\sigma^2 = \sigma_{0}^{2} \left (1 + \frac{2\tau}{\delta t} \right )
\end{equation}
where   $ \sigma_{0}$ is the standard deviation treating all
data  as they were statistically independent
 and  $ \sigma$ is the actual statistical
uncertainty. This variance inflation correctly takes   into account the
correlations of the MC data.

It is not a simple matter to access the actual error  in a finite time series of correlated   data.
Here we employed the moving block bootstrap (MBB) method\cite{mbb} which
exploits resampling techniques.
Within the MBB scheme a block of observations is defined by its length and
 by its starting point in the series.    For instance, $Q_i = \{q_i, q_{i+1},\ldots,q_{i+l}\}$
defines the i${th}$ block of $l$ observations. A MBB sample is
then obtained by: (i) randomly drawing with replacement from the set of all
possible overlapping blocks of size $l$; (ii) concatenating the selected blocks
forming a replicated series.
Each set of replicated data obtained in this way yields one estimate for the sample mean.
The drawing is repeated many times  and
the block size dependent error is approximated by the standard deviation of the bootstrap
generated mean
values.

It can be shown that in the case of arithmetic mean, $\sigma^2$ can be calculated exactly
without resampling\cite{jacknife}. For a series with $n$ observations, $q_t$, and $k$ blocks of
size   $l$,  $\sigma^2$ is given by\cite{mbb}
\begin{equation}
\sigma^2 = \frac{1}{kn} \sum_{j=0}^{n-1} \left [ \frac{1}{l}
 \sum_{t=1}^{l} (q_{j+t}-\langle q \rangle) \right ]^2.
\end{equation}

The behavior of the ratio $\sigma^2 / \sigma_{0}^{2}$
is illustrated in figure~7 for the susceptibility.
 The error increases with the block size until it becomes roughly size independent for
block length large enough. The maximum value reached by    error corresponds to the
actual standard error of the mean.

The underlying  idea of the MBB method is that if the block length is large enough,
observations belonging to different blocks are nearly independent, while the correlation
 present in observations forming each block is retained.

 Having an estimate to   $\sigma^2 / \sigma_{0}^{2}$, Eq. 13 can be employed
to extract $\tau_{int}$. The above outlined procedure was applied for the data
of the susceptibility and energy of all lattices. Good agreement was achieved
between the estimates of $\tau_{int}$ obtained from MBB and by applying
directly Eq. 12 for small lattices.

The computed  equilibrium relaxation times from both, susceptibility
 and energy, are plotted in figure~8 as obtained from lattices of  size $L=16, 24, 28, \ldots, 48$.
Notice that, although $\tau_{int}$ is quite smaller for the susceptibility, both
exhibit the same power-law size dependence. A linear best fit for  the energy and susceptibility
data
yields $ 3.0 \pm 0.1$ for the regression coefficient. Therefore, the equilibrium relaxation time
scales as $\tau_{int}\propto L^{3.0 \pm 0.1}$. This means that, within the Handscomb
 dynamics, the number of Monte Carlo steps per site required to generate 
uncorrelated diagram configurations at criticality is roughly size independent.

\section{conclusions}

In summary, we performed Monte Carlo simulations of the $S=1/2$ Heisenberg 
ferromagnet on the simple cubic lattice to investigate the critical relaxation 
of the Handscomb quantum Monte Carlo method which samples the space of permutation
 operators appearing in the series expansion of the partition function. Precise 
estimates of the critical temperature and exponents $\gamma/\nu$ and $\beta /\nu$ 
were obtained from a phenomenological renormalization group analysis of data from 
the susceptibility and order parameter. At the critical temperature we measured
 the equilibrium  relaxation time  from the time-displaced correlation
 functions of the susceptibility and energy (small lattices only).
For the largest lattices ($L \ge 32$) $\tau_{int}$ was estimated
 through the moving block bootstrap technique. From either susceptibility or energy
 we obtained that, at criticality,
 the number of Monte Carlo steps (sampled permutation sequences) required to generate
 uncorrelated equilibrium diagram configurations scales with the system's volume. 

Some care must be taken when estimating the efficiency of the Handscomb method 
and comparing it with other Monte Carlo prescriptions. Firstly, the phase space 
sampled within the Handscomb method is not related to any physical space. Therefore,
 there is no direct relation between the time scales of the Handscomb and the traditional
 spin-flip dynamics. However, a crude estimate can be drawn by considering that during an
 elementary Monte Carlo step of the Handscomb dynamics the sites belonging to a
 particular cycle of permutations have their status updated. The average  number of sites involved 
in a single Monte Carlo step is then proportional to
 $\frac{1}{N}\left< a_i^2\right> \sim \chi\sim L^{\gamma/\nu}$. Within this 
reasoning, the average fraction of sites updated in a MCS scales as 
$L^{\gamma/\nu}/L^d$. Therefore, a time scale which would correspond to 
a lattice sweep in spin-flip dynamics would be $\tau_0\sim L^{d-\gamma/\nu}$.
 In units of this time scale the relaxation time scales as $\tau_{int}\sim \tau_0L^z$, with 
$z =2\pm 0.1$, which is quite similar to the value of $z$ found for the decoupled cell
Quantum Monte Carlo
and the classical Metropolis dynamics. Although the Handscomb dynamics depicts some
characteristics of the classical spin-flip cluster dynamics it has not a similar
 effect on dealing with the critical slowing down.  

It is relevant to mention here that the present Handscomb prescription, which inserts
 or removes transposition operators at the extremes of the permutation sequence, is 
the one that provide the simplest algorithm to control the dynamics in the permutation
 phase space.  A natural generalization is to insert and remove operators at random 
locations within the sequence. This would drastically change its cycle structure 
 with all sites being able to have their status updated on a single step. It would 
be valuable to estimate the efficiency of such relaxational dynamics at criticality as well as
 that of other generalizations of the Handscomb prescription as applied to antiferromagnet 
and large spin models.

\section{acknowledgments}

We are indebted to D.P. Landau for his suggestions and critical reading of the
manuscript. This work was partially supported by CNPq and CAPES
 (Brazilian research agencies). MLL
 would like to thank the hospitality of the Physics Department at Universidade Federal
 de Pernambuco during the Summer School 2000 where this work was partially developed.

\newpage

\section*{FIGURE CAPTIONS}

Fig.1 -  The susceptibility per spin as a function of temperature for $L=16, 24$ and $32$
(from below). Due to an intrinsic symmetry of the Handscomb dynamics, the susceptibility
 equals the magnetization second moment below $T_c$. The errors are much smaller
 than the size of the symbols.

Fig.2 -  The auxiliary functions $g_{\chi}(T,L,L')$ for the scaling of  
susceptibility data. The renormalizations were performed from $L=24$ to $L'=16$
(circles); $L=32$ to $L'=16$ (squares); $L=40$ to $L'=16$ (diamonds) and from
$L=40$ to $L'=24$ (triangles). Typical error bars are shown.
The solid lines are the results from  renormalizations of the best fits of our original
 susceptibility data. These have a common point from which we estimate
$T_c=1.677\pm 0.001$ and
$\gamma/\nu = 1.98\pm 0.01$.

Fig.3 - The average size of the largest cycle (normalized the the total number of sites) $\psi$
 as a function of temperature for
$L=16, 24$ and $48$. At high temperatures all cycles are small indicating
no long range order and $\Psi$ vanishes. With lowering $T$,  the onset of the  ferromagnetic
order makes itself felt around $k_BT_c/J \approx 1.68$, and $\psi$ start to grow until saturation.
At criticality,  $\psi$ shows power law size dependence.

Fig.4 - The auxiliary functions $g_{\psi}(T,L,L')$ for the scaling of
 the order parameter data. The renormalizations were performed from $L=24$ to $L'=16$
(circles); $L=40$ to $L'=16$ (squares); $L=48$ to $L'=16$ (triangles up) and
from $L=48$ to $L'=24$ (triangles down). Typical error bars are shown.
The solid lines are the results from
 renormalizations of the best fit of our original order parameter data. From the
 interception of these functions
computed for all possible renormalizations with lattice sizes $L=16,
 24, 32, 40$ and $48$ we estimate $k_BT_C/J = 1.6778 \pm 0.0002$ and
 $\beta /\nu = 0.512 \pm 0.002$.

Fig.5 - An equilibrium trace of the susceptibility $\chi$ and energy $E$ at criticality.
Local quantities, as energy, are more time correlated than non-local ones due to
the cluster nature of the Handscomb Monte Carlo.

Fig.6 - The time-displaced equilibrium correlation function of the susceptibility and energy
at criticality for several lattice sizes.

Fig.7 - Moving block bootstrap estimates of the standard errors of the
susceptibility as a function of the block length $l$ at criticality for
several lattice sizes. The lines are guides to the eye.

Fig.8 - The equilibrium  relaxation time versus linear size $L$
 for susceptibility and energy. The error in our estimates of $\tau_{int}$ is
  around 2\%. Though,  $\tau_{int}$ is much smaller for the susceptibility,
 both quantities scale the same way. The microscopic time scale used is $2000$ MCS.

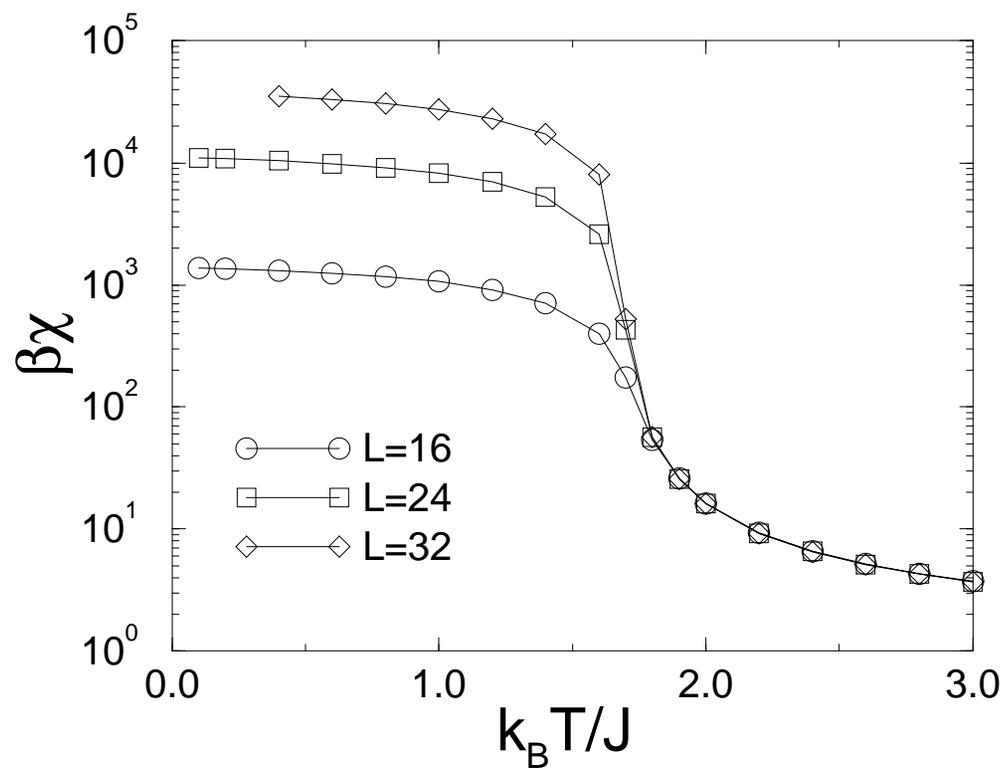
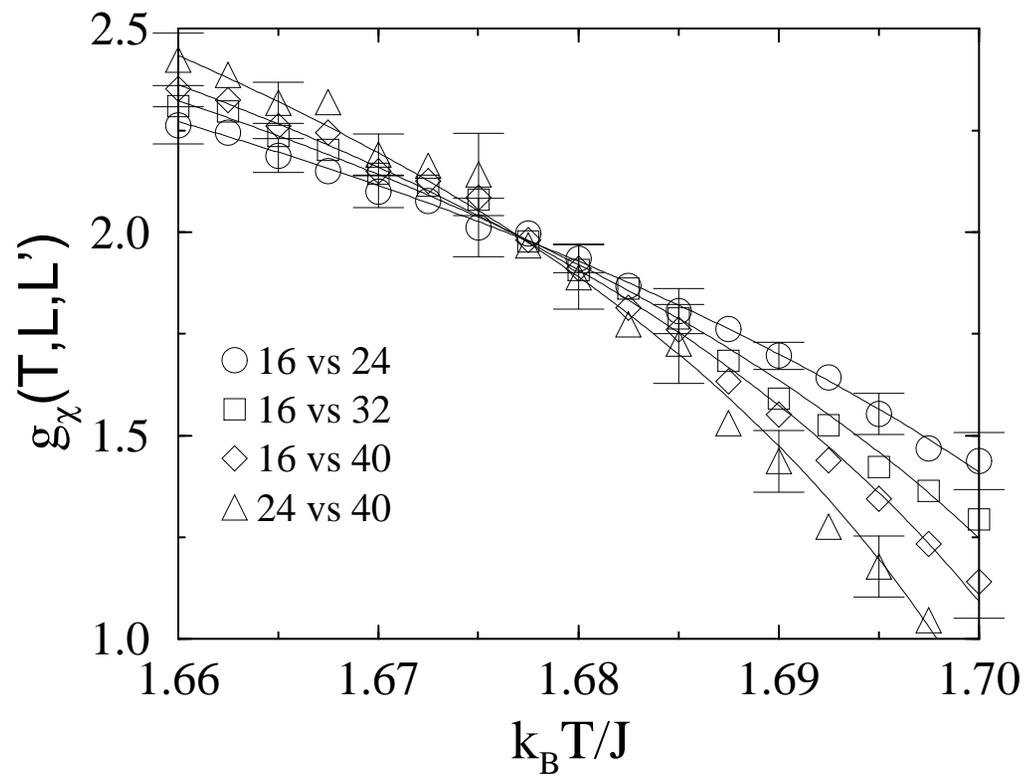
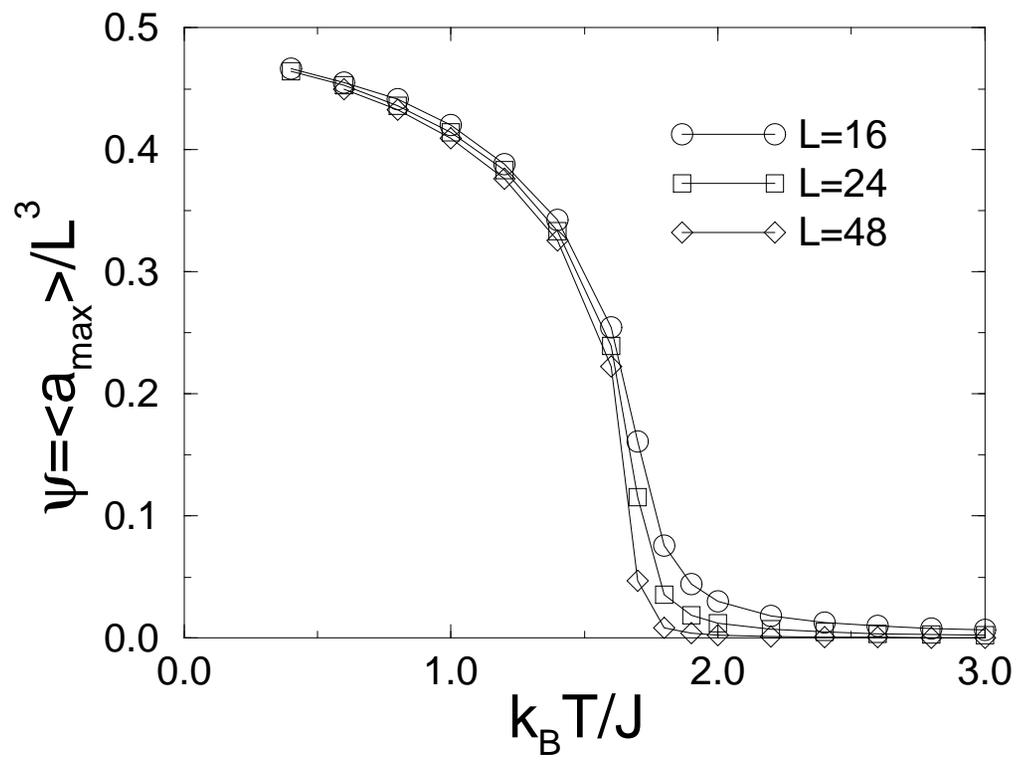
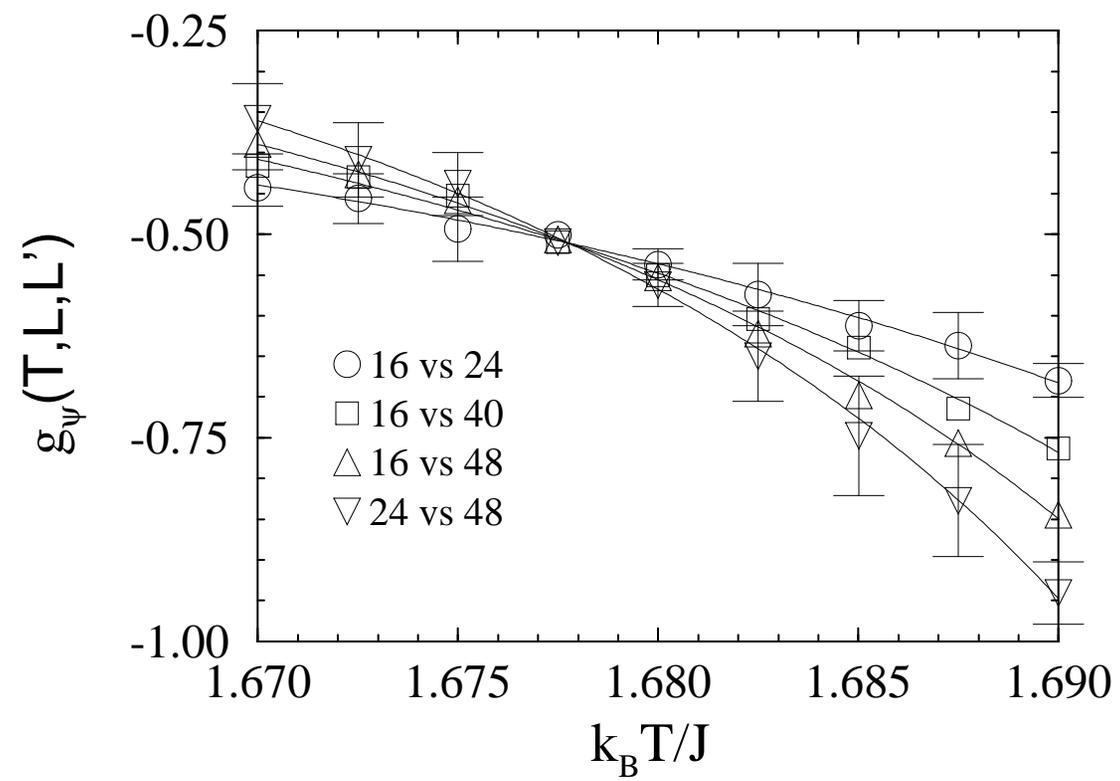
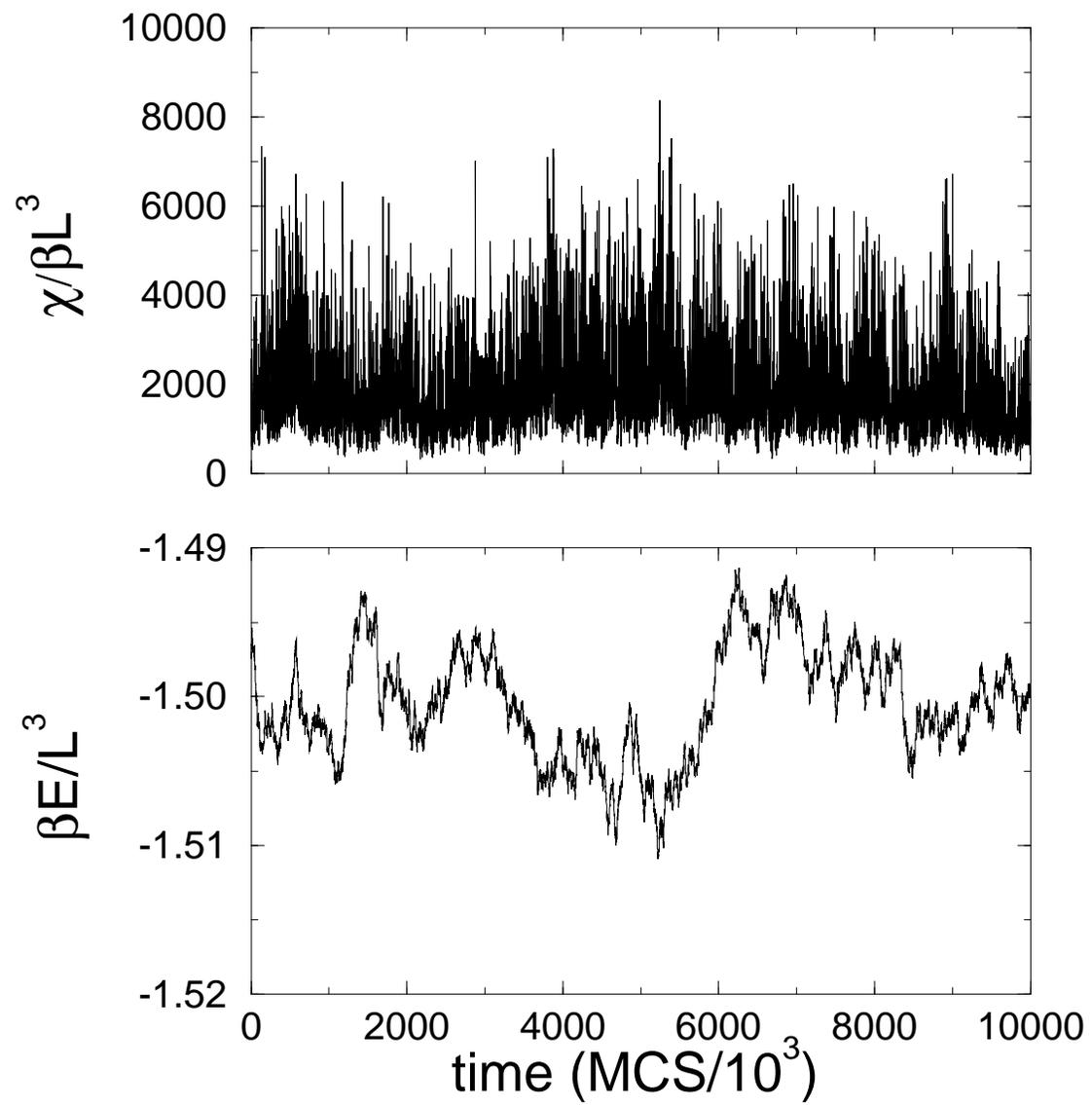
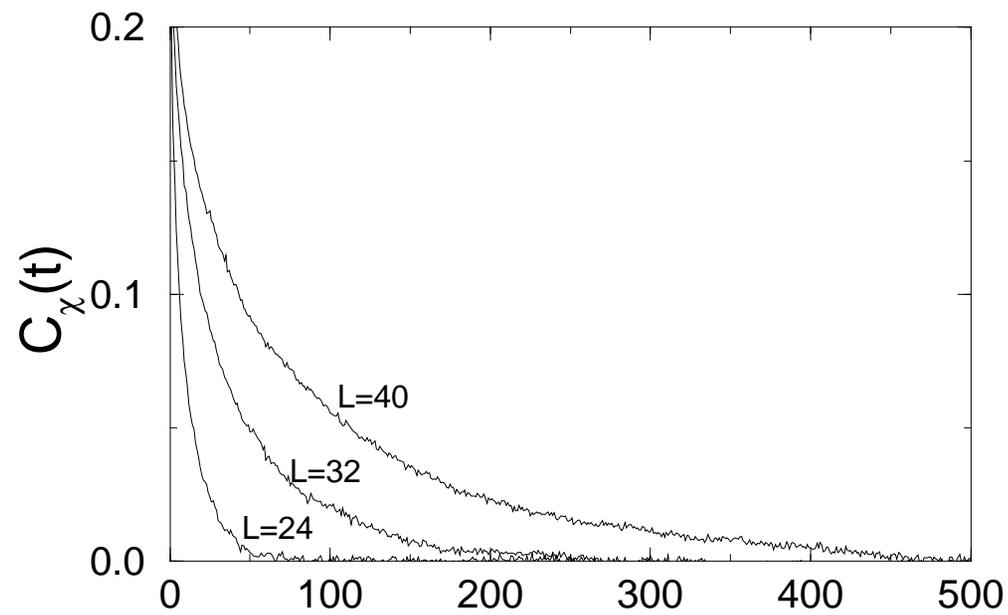
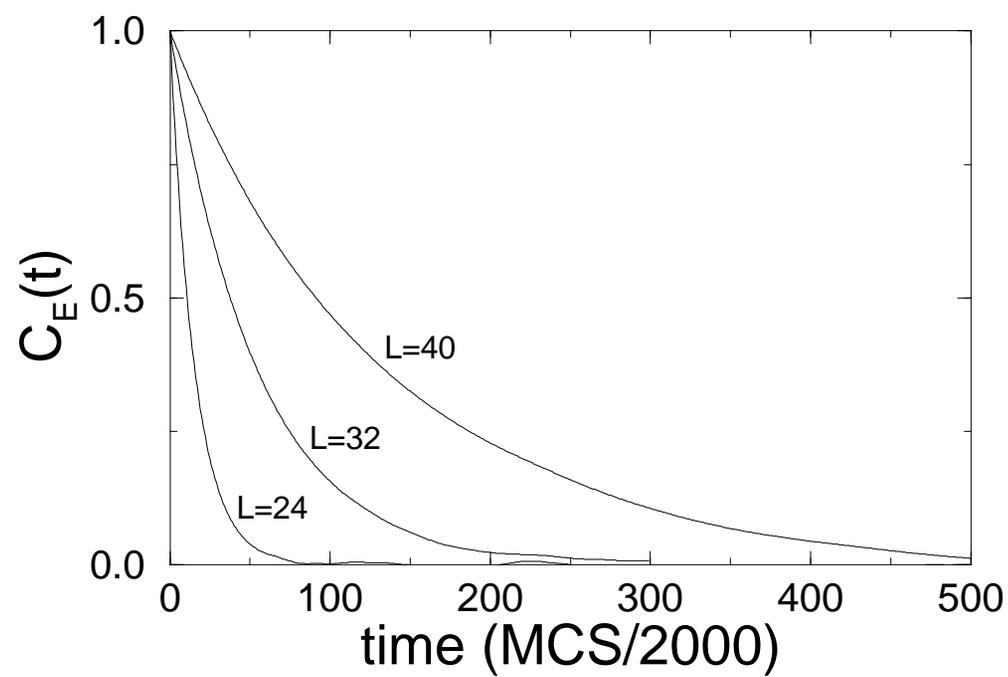
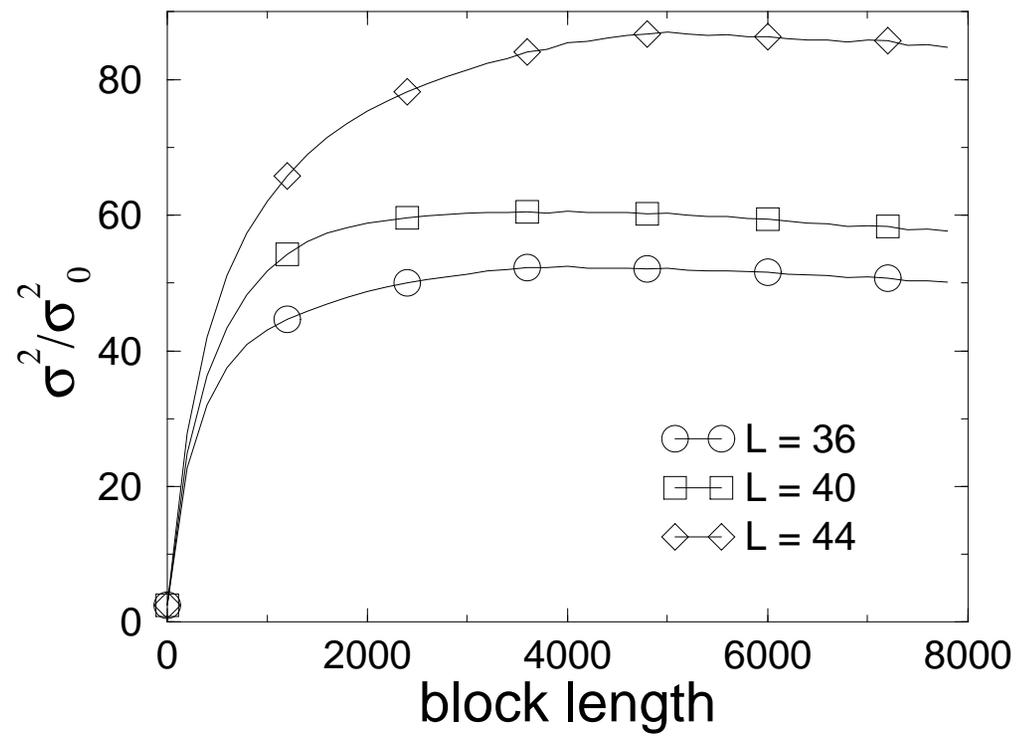
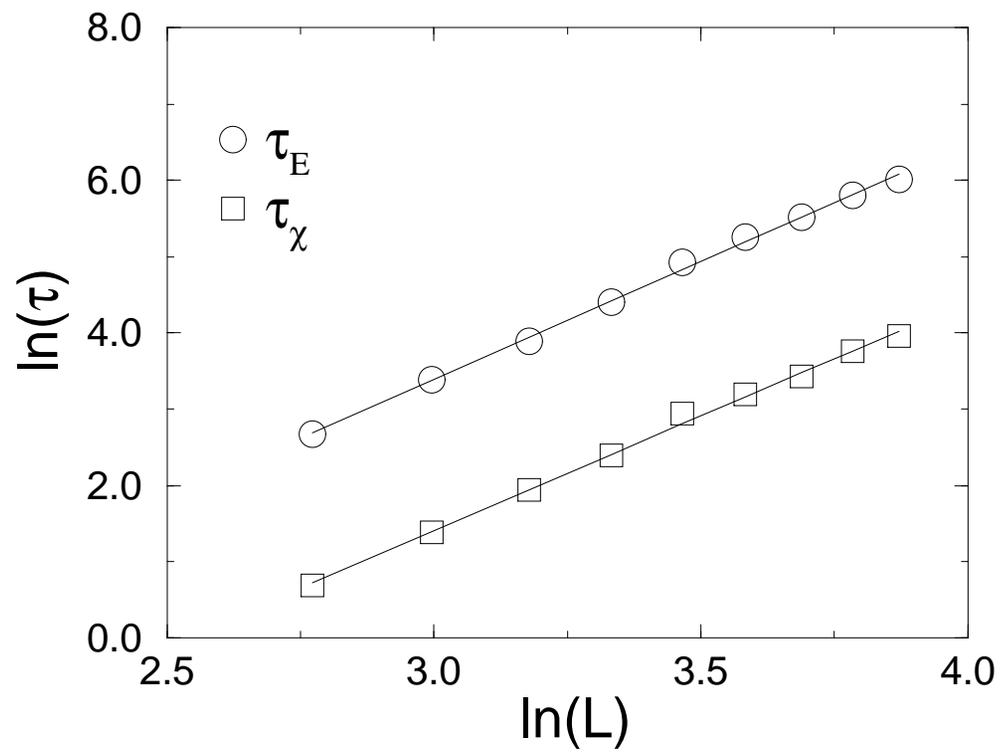
\end{document}